\documentclass{aa}  

\usepackage{graphicx}
\usepackage{txfonts}
\usepackage{color}

\begin{document}

   \title{Are LGRBs biased tracers of star formation? Clues from the host galaxies of the {\it Swift}/BAT6 complete sample of LGRBs. \\
   I: Stellar mass at  $z<1$  }


   \author{S.D. Vergani
          \inst{1,2}
          \and
          R. Salvaterra
          \inst{3}
          \and 
          J. Japelj
          \inst{4} 
          \and
          E. Le Floc'h
          \inst{5}
          \and
          P. D'Avanzo
          \inst{2}
          \and
          A. Fernandez-Soto
          \inst{6,7}
          \and
          T. Kr\"uhler
          \inst{8}
          \and
          A. Melandri
          \inst{2}
          \and
          S. Boissier
          \inst{9}
          \and
          S. Covino
          \inst{2}
          \and
          M. Puech
          \inst{1}
          \and
          J. Greiner
          \inst{10}
          \and
          L.K. Hunt
          \inst{11}
          \and
          D. Perley
          \inst{12}
          \and
          P. Petitjean
          \inst{13}
          \and
          T. Vinci
          \inst{14}
          \and
          F. Hammer
          \inst{1}
        \and
          A. Levan
          \inst{15}
          \and
          F. Mannucci
          \inst{11}
          \and
          S. Campana
           \inst{2}
           \and
           H. Flores
          \inst{1} 
          \and
          A. Gomboc
          \inst{4}    
        \and
          G. Tagliaferri
          \inst{2}
          }

   \institute{GEPI-Observatoire de Paris Meudon. 5 Place Jules Jannsen, F-92195, Meudon, France\\
               \email{susanna.vergani@obspm.fr}
         \and
         INAF, Osservatorio Astronomico di Brera, via E. Bianchi 46, 23807 Merate, Italy
         \and
         INAF, IASF Milano, via E. Bassini 15, I-20133 Milano, Italy
         \and
         Faculty of Mathematics and Physics, University of Ljubljana, Jadranska ulica 19, SI-1000 Ljubljana, Slovenia
         \and
         Laboratoire AIM, CEA/DSM/IRFU, CNRS, Universit\'e Paris-Diderot,
91190 Gif, France
         \and
         Instituto de F\'isica de Cantabria (CSIC-UC), E-39005 Santander, Spain
         \and
         Unidad Asociada Observatori Astron\'omic (IFCA - Universitat de Val\'encia), Valencia, Spain
         \and
         European Southern Observatory, Alonso de C\'ordova 3107, Vitacura, Casilla 19001, Santiago 19, Chile
         \and
         Aix Marseille Universit\'e, CNRS, LAM (Laboratoire d'Astrophysique de Marseille) UMR 7326, 13388, Marseille, France
         \and
         Max-Planck-Institut f\"ur extraterrestrische Physik, Giessenbachstrasse 1, 85748, Garching, Germany
         \and
         INAF - Osservatorio Astrofisico di Arcetri, Largo E. Fermi 5, I-50125 Firenze, Italy
         \and
         Department of Astronomy, California Institute of Technology, MC 249-17, 1200 East California Blvd, Pasadena, CA 91125, USA
         \and
         University Paris 6, Institut d'Astrophysique de Paris, UMR7095 CNRS, 98bis Boulevard Arago, F-75014, Paris, France
         \and
         LULI, Ecole Polytechnique, CNRS, CEA, UPMC, 91128 Palaiseau, France
         \and
         Department of Physics, University of Warwick, Coventry, CV4 7AL, UK
             }

   \date{Received ???}

 
  \abstract{}
  {Long gamma-ray bursts (LGRBs) are associated with massive stars and are therefore linked to star formation. 
  However, the conditions needed for the progenitor stars to produce LGRBs can affect the relation between the LGRB rate and star formation.
By using the power of a complete LGRB sample, our long-term aim is to understand whether such a bias exists and, if it does, what its origin is.  
}
  {To reach our goal we use the {\it Swift}/BAT6 complete sample of LGRBs. In this first paper, we build the spectral energy distribution (SED) of the 14 $z<1$ host galaxies of the BAT6 LGRB sample and determine their stellar masses (M\,$_{\star}$) from SED fitting. To investigate the presence of a bias in the LGRB-star formation relation we compare the stellar mass distribution of the LGRB host galaxies (i) with star-forming galaxies observed in deep surveys (UltraVISTA) within the same redshift limit, (ii) with semi-analytical models of the $z<1$ star-forming galaxy population, and (iii) with dedicated numerical simulations of LGRB hosts having different metallicity thresholds for the progenitor star environment.
}
  {We find that at $z<1$, LGRBs tend to avoid massive galaxies and are very powerful for selecting a population of faint low-mass star-forming galaxies, partly below the completeness limits of galaxy surveys. The stellar mass distribution of the hosts is not consistent with that of the UltraVISTA star-forming galaxies weighted by their star formation rate (SFR). This implies that, at least at $z < 1$, LGRBs are not unbiased tracers of star formation. To make the two distributions consistent, a much steeper faint end of the mass function would be required or a very shallow SFR-mass relation for the low-mass galaxy population. The comparison with the GRB host galaxy simulations indicates that, to reproduce the stellar mass distribution, a metallicity threshold of the order of Z$_{th}$=0.3-0.5 Z$_{\odot}$ is necessary to form a LGRB. Models without a metallicity threshold or with an extreme threshold of $Z_{th}=0.1Z_{\odot}$ are excluded at z<1. Under a very basic assumption, we estimate that the LGRB rate can directly trace the SFR starting from $z\sim4$ and above. 
}
  {GRB hosts at $z<1$ have lower luminosities and stellar masses than expected if LGRBs were unbiased star formation tracers. The use of the BAT6 complete sample keeps this result from being affected by possible biases that could have influenced past results based on incomplete samples. The preference for low metallicities (Z$\lesssim 0.5$Z$_{\odot}$) inferred by the comparison with the simulations can be a consequence of the particular conditions needed for the progenitor star to produce a GRB. }{} 

   \keywords{Gamma-ray burst: general - Galaxies: photometry - Galaxies: star formation
               }

   \titlerunning{Are LGRBs biased tracers of star formation? Clues from the BAT6 sample }

   \maketitle

%

\section{Introduction}

In recent years, great progress in understanding cosmic star formation has come
through studying populations of star-forming galaxies selected through different techniques. However, each sample
suffers from uncertainties arising from dust extinction, flux-limited samples, and incomplete redshift determinations. 
Here is where long gamma-ray bursts (LGRBs, prompt emission duration longer
than 2 seconds) can play an important role. 

GRBs are short, intense bursts of gamma-rays that during seconds to minutes outshine all other sources of gamma-ray emission in the sky.
Following the prompt gamma-ray emission an `afterglow' emission in the range from X-ray to radio wavelengths
persists up to months after the initial burst. Uniquely, GRBs allow galaxies to be selected independently of
their brightnesses at any wavelength. The association of LGRBs with the explosion of broad-line type Ic SNe (see \citealt{Hjorth2011} for a review), 
hence with the death of massive stars, makes this class of GRBs a unique tool for understand the evolution of
star formation and galaxies, complementary to current galaxy surveys, 
up to the highest redshifts ($z=8.2$; \citealt{Salvaterra2009,Tanvir2009}; $z\sim9$; \citealt{Cucchiara2011}).

Indeed, the observations of the $z>5$ LGRB host galaxies resulted in very deep limits on their
magnitudes, which are as faint as 30.3(AB) in H-band \citep{Tanvir2012,Basa2012}. This is consistent with the results of numerical simulations showing that LGRBs pinpoint 
star-forming galaxies belonging to the faint end of the luminosity function. Such galaxies form the bulk of the
first galaxies population that should significantly contribute to re-ionisation \citep{Salvaterra2011}.

Several studies have targeted LGRB host galaxies (e.g. \citealt{Lefloch2003,Chris2004,Fruchter2006,Lefloch2006,Savaglio2009,Castro-Ceron2010,Svensson2010,Kruhler2015}). LGRB hosts at $z<1$ are in general found to be sub-luminous, blue, moderately
star-forming (SFR$\sim1$\,M$_{\odot}$\,yr$^{-1}$), and with stellar masses M\,$_{\star}\sim1-5\times10^9$\,M$_{\odot}$ (but see e.g. \citealt{Kruhler2011,Kruhler2015}). 
The observed luminosity distribution of LGRB hosts is broad, indicating
that a considerable amount of star formation takes place in low luminosity galaxies.
\cite{Han2010}, \cite{Levesque2010}, \cite{Graham2013}, \cite{Wang2014} show that LGRB host galaxies (at least at $z<1$) have a lower metallicity
than do other galaxies at similar masses, tracing an environment that is less chemically evolved than the
galaxies selected by other techniques. On the other hand, at $z<1$, LGRB hosts are
found to be consistent with the fundamental metallicity relation (FMR; \citealt{Mannucci2010,Mannucci2011}), 
disfavouring the very low metallicity thresholds expected from the LGRB theoretical models, which would
have a strong impact on the properties of the galaxy population hosting the LGRB event. 

The host galaxies of LGRBs have also been studied by numerical simulations or semi-analytical models (e.g.: \citealt{Nuza2007, Chisari2010,Mao2010,Artale2011,Niino2011,Elliott2012,Trenti2014}).
\cite{Campisi2011} studied the mass-metallicity relation and the FMR of LGRB host galaxies comparing 18 LGRB host galaxies with a catalogue of simulated galaxies
constructed combining high-resolution N-body simulations with a semi-analytic model of galaxy formation. They found that a very low-metallicity threshold (Z\,$<0.3$Z${_\odot}$) is not required to reproduce either relations, explaining the observed low metallicities as a consequence of LGRBs exploding in low-mass and strongly star-forming galaxies.

With a new approach, \cite{Boissier2013} 
tried to determine the GRB efficiency (i.e. the GRB rate - SFR ratio) and its causes more quantitatively by studying the distribution of LGRB host properties with respect to the
star-forming galaxy population,
so as to still be able to properly use LGRBs
as SFR tracers.

The picture seems to become more complex when the host galaxies of the so-called {\it \emph{dark}} LGRBs (see \citealt{Jakobsson2004,van-der-Horst2009}, for a definition) are considered, and when the studies are extended to higher redshift and/or longer wavelength domains.
\cite{Kruhler2011}, \cite{Perley2013}, and \cite{Hunt2014} found that on average \emph{{\it \emph{dark}}} LGRBs extend the LGRB host galaxy population to more massive galaxies with a higher SFR. For some of the few host galaxies detected in the far-infrared, sub-millimetric, and radio domains (e.g. \citealt{Berger2003,Michaowski2008,Michaowski2012,Hunt2014,Schady2014,Symeonidis2014,Perley2015b}), very high SFR values have been found, up to $10^3$\,M$_{\odot}$\,yr$^{-1}$. 

All the observational studies cited above are based on different, heterogeneous samples of LGRB host galaxies, which are far from being complete and which are likely to be biased, therefore they can obtain different results on LGRB host galaxies properties (e.g. Perley et al. 2013 and Hunt et al. 2014 reach different conclusions about GRBs as SFR tracers). Nonetheless, there is some agreement on the fact that, at z < 1, LGRBs seem to prefer lower mass systems relative to a purely SFR-selected galaxy sample.
To have an unbiased picture of the properties of LGRB hosts,
complete samples of LGRBs are necessary. 

Three complete samples of GRBs have become available between 2011 and 2012: the GROND\footnote{\url{http://www.mpe.mpg.de/~jcg/GROND/}} \citep{Greiner2008} sample \citep{Greiner2011}, the
BAT6 sample presented in \cite{Salvaterra2012}, and the TOUGH sample \citep{Hjorth2012}. Furthermore, \cite{Perley2015,Perley2015a} present a study based on Spitzer observation of the unbiased Swift Host Galaxy Legacy (SHOALS) Survey. 
In this paper we focus on the BAT6 sample.

The BAT6 LGRBs are selected only on the basis of the brightness
of their gamma-ray prompt emission (Swift/BAT peak 
flux $P\ge 2.6$ ph\,s$^{-1}$\,cm$^{-2}$), and on favourable observing
conditions (practically not introducing any bias, see \citealt{Jakobsson2006}). 
The cut on the brightness of the gamma-ray emission is needed to guarantee the completeness. In fact, getting closer to the BAT detection threshold, {\it Swift} starts to miss some GRBs.
The sample consists of 58 LGRBs, and it is complete in redshift
at 95\%. We note that the detection of the optical afterglow is not required in the construction of the BAT6
sample, therefore dark bursts are represented well within the sample (see \citealt{Melandri2012}). This sample
offers the unique opportunity to study the population of LGRBs in a statistically unbiased way. In particular,
as the brightness of the gamma-ray prompt emission is the only burst-depending parameter for selecting the LGRB sample, the BAT6 allows the selection of a complete sample of LGRB host galaxies independently
of their fluxes or colours.

Our project is to study the properties of the host galaxies of the $z<1$ LGRBs of the BAT6 with the aim of investigating the relation between LGRB and star formation.
In this paper we present the study of the stellar masses. 
We restrict ourselves to LGRBs at $z<1$ because i) of
the wealth of data available in this redshift range, ii) of the larger differences expected in this
redshift range on theoretical grounds between the LGRB hosts and the normal field populations, and iii) of the range where the study of the properties
of the galaxies can be carried out by optical/near-infrared instruments. 
It is not possible at the moment to maintain the same completeness level in the galaxy property determination at higher redshift for our sample.

In section 2 we present the host galaxy sample and the data used. The K luminosity and the stellar masses are studied in section 3 and compared to those of the UltraVISTA survey in section 4. Application of the Boissier's method to our data is presented in section 5, whereas in section 6 we compare the stellar mass distribution with the LGRB host galaxy simulations. Discussion and conclusions are carried out in sections 7 and 8, respectively.


\section{The sample of LGRB host galaxies }
We selected the $z<1$ LGRB in the BAT6 sample, corresponding to 14 objects. To determine the stellar masses of the host galaxies of the selected objects, we collected the photometry published in the literature and all the available photometric data. 
In addition we obtained new data to complete the spectral energy distributions (SEDs) when needed. We also checked for {\it Spitzer} observations and found that 12 of 14 of the hosts have IRAC1 or IRAC2 observations. More than half of the hosts in the sample are detected, and upper limits can be obtained for the remaining observed hosts, allowing the coverage of the rest-frame K band. 

All the magnitudes collected for each galaxies can be found in Table \ref{mag}.
We give here a brief description of the newly determined photometry, both from our new observational programmes and from the public unpublished data. These magnitudes were corrected for the Milky Way extinction according to the values of \cite{Schlafly2011}, assuming RV = 3.08. More details on all the photometry used for each host galaxy can be found in the appendix.

\subsection{New GROND data and photometry}
\label{GROND}
We obtained GROND data that was not published before for the host galaxies of GRB\,080916A, GRB\,081007 and GRB\,091018, in all the GROND 7 filters.
All data were calibrated against the SDSS or 2MASS.
Standard aperture photometry was used to derive the host flux with an appropriate aperture correction to get total fluxes.
The magnitudes are based on a stack of different dithered images (12-14 in $g\,r\,i\,z$, 360-480 in $JHK$) and were only those images that fulfil a certain set of criteria in terms image quality are used, as explained in details in \cite{Kruhler2011}.
The errors include both the statistical errors of the detection and the error in the photometric calibration.

\subsection{New GTC data and photometry}
\label{GTC}
GTC data were obtained with OSIRIS for the host galaxies of GRB\,050525 ($g$ and $i$ bands) and GRB\,071112C ($z$ band) under the programme GTC31-13B (PI: A. Fernandez-Soto).
The data were reduced using standard procedures with tools provided by the ESO-Eclipse package \citep{Devillard1997}. Photometry was carried out with SExtractor (v2.8.6, \citealt{Bertin1996}). Instrumental zero points were measured by observing standard stars SA112-805 and SA104-428 \citep{Smith2002} in the case of 050525A and 071112C hosts, respectively. Standard aperture photometry with an appropriate aperture correction was applied to derive the host galaxy magnitudes. Final errors include both statistical errors and the uncertainties of the photometric calibration.

\subsection{New VLT/HAWK-I data and photometry}
\label{HAWK}
New VLT/HAWK-I data were obtained for the host galaxies of GRB\,061021, GRB\,091018, GRB\,080916A (Programme ID 092.D-0305, PI: S.D. Vergani), and GRB\,081007 (Programme ID 092.A-0231, PI: T. Kruhler). The data were obtained with the $J$ and $K$ filters for all the host galaxies except for GRB\,091018 for which only the $K$ filter has been used.

HAWK-I images were reduced using the ESO HAWK-I pipeline v1.8.12\footnote{\url{http://www.eso.org/sci/software/pipelines/}} following the monolithic reduction cascade\footnote{ftp://ftp.eso.org/pub/dfs/pipelines/hawki/hawki-pipeline-manual-1.9.pdf}, which includes background subtraction, flat field and illumination correction, sky subtraction, astrometric offset refinement, stacking of the images and distortion correction. Photometry was carried out with SExtractor (v2.8.6, \citealt{Bertin1996}). Standard aperture photometry with an appropriate aperture correction has been applied to derive the host galaxy magnitudes. If a host is not detected in the image, we report 3$\sigma$ upper limits. Magnitudes have been calibrated against several unsaturated 2MASS stars in the field of view in the science frames. Final errors include both statistical errors as well as the uncertainties of the photometric calibration.

\subsection{New TNG data and photometry}
\label{TNG}
New TNG photometry has been obtained for the hosts of GRB\,080430 (OPTICON ID: OPT13B\_71; PI S.D. Vergani; $g i z$ filters). We used also some public but unpublished data (programme ID: A22\_TAC107, PI: D. Malesani), retrieved from the TNG archive (\url{http://ia2.oats.inaf.it/index.php/tngarchive/tng}) for the host galaxies of GRB060912A ($UBVIJ$ filters), GRB061021 ($B$ and $V$ filters) , GRB071112C (program ID: A17\_TAC9, PI: C. Guidorzi; $I$ filter) 

Image reduction was carried out following standard procedures: subtraction of an averaged bias frame and division by a normalised flat frame. The photometric calibration was achieved by observing Landolt standard fields. Aperture photometry was performed using the SExtractor package \citep{Bertin1996}. Astrometric solutions were computed against the USNO-B1.0 catalogue\footnote{\url{http://www.nofs.navy.mil/data/fhcpix}}.

\subsection{New Gemini photometry}
\label{Gemini}
For the host galaxy of GRB\,071112C,
we used public $R$-band Gemini data that has not been published previously, retrieved from the Gemini archive (\url{http://www2.cadc-ccda.hia-iha.nrc-cnrc.gc.ca/en/gsa/}; program ID:\,GN-2008B-Q-112 PI:\,A. Levan).
Image reduction was carried out following the standard procedures described in Sect. \ref{TNG}.

\subsection{New HST photometry}
\label{HST}
We retrieved from the HST archive (\url{https://archive.stsci.edu/hst/search.php}) the F160W public HST data (not published before), for the host galaxies of GRB\,071112C, GRB\,080430, GRB\,080916A, and GRB\,081007 (programme ID: 12307, PI: A. Levan). 
HST/WFC3 observations were reduced in the standard HST fashion. Photometry was carried out with SExtractor (v2.8.6, \citealt{Bertin1996}) and calibrated with the official WFC3 instrument zero points\footnote{\url{http://www.stsci.edu/hst/wfc3/phot\_zp\_lbn}}. Standard aperture photometry with an appropriate aperture correction has been applied to derive the host galaxy magnitudes.

\subsection{New {\it Spitzer} photometry}
We use public {\it Spitzer} data not previously published, retrieved from the {\it Spitzer} archive (\url{http://irsa.ipac.caltech.edu/Missions/spitzer.html}), for all the host in the sample except for GRB\,050416 and GRB\,080319B. The data come from the different observing programmes of D.\,Perley (see the Appendix for details) except for the host of GRB\,050525 (programme ID:\,3653, PI: P.M. Garnavich).
The IRAC fluxes were determined using the PBCD (post-basic calibrated data) and applying aperture photometry (1.9\arcsec\,radius) following the prescriptions of \cite{Sanders2007}.

Given the lower spatial resolution of the IRAC data, special attention was devoted to addressing possible causes of confusion. Significant flux contamination and blending was found for only one object in our sample (GRB\,060912A), which lies $\sim$11\arcsec away from a much brighter nearby galaxy. The contribution of the latter within the apertures used for the GRB host photometry (sky + source) was estimated pixel by pixel by fitting its surface brightness with a Sersic profile model, which was subsequently subtracted from the IRAC image before estimating the GRB host photometry.

\begin{sidewaystable*}[h!]
\caption{Observed {\it AB} magnitudes (corrected by the Milky Way extinction) of the LGRB host galaxies of the BAT6 sample at $z<1$. New photometry is in bold.}             
\label{mag}      
\centering     
\small 
\begin{tabular}{l c c c c c c c c c c }     
\hline\hline  
 \multicolumn{11}{c}{Visible}  \\
\hline                       
Host galaxy & $z$ & $U$ & $B$ & $V$ & $R$ & $I$ & $g$ & $r$ & $i$ & $z$ \\    
\hline     
\object{GRB\,050416}& 0.653 &  $ $ & $ $ & $ $ & $23.19\pm0.02$ & $23.05\pm0.10$ & $24.00\pm0.03$ & $ $ & $ $ & $ $\\
\object{GRB\,050525}& 0.606&  $ $ & $ $ & $ $ & $25.96\pm0.34$ & $ $ & ${\bf 27.7\pm0.4}$ & $ $ & ${\bf 26.9\pm0.3}$ & $ $   \\
\object{ GRB\,060614}& 0.125 &  $23.79\pm0.23$ & $23.45\pm0.15$ & $22.71\pm0.07$ & $22.69\pm0.06$ & $22.18\pm0.07$ & $ $ & $$ & $$ & $$     \\     
\object{GRB\,060912}& 0.937 &  ${\bf 23.77\pm0.14}$ & ${\bf 23.26\pm0.11}$ & ${\bf 23.35\pm0.14}$ & $22.78\pm0.04$ & ${\bf 22.80\pm0.13}$ & $$ & $$ & $$ & $$ \\
\object{GRB\,061021}& 0.346 &  $$ & ${\bf 25.47\pm0.36}$ & ${\bf 25.22\pm0.32}$ & $24.53\pm0.06$ & $$ & $$ & $$ & $$ & $$ \\
\object{GRB\,071112C}& 0.823 &  $$ & $$ & $$ & ${\bf 25.21\pm0.12}$ & ${\bf 24.54\pm0.13}$ & $$ & $$ & $$ & ${\bf 24.05\pm0.14}$ \\
\object{GRB\,080319B}& 0.937 &  $$ & $$ & $$ & $$ & $$ & $26.81\pm0.14$ & $26.96\pm0.13$ & $26.17\pm0.15$ & $$\\
\object{GRB\,080430}& 0.767 &  $$ & $$ & $$ & $$ & $$ & ${\bf 25.21\pm0.19}$ & $24.26\pm0.10$ & ${\bf 24.43\pm0.22}$ & ${\bf 24.05\pm0.15}$ \\
\object{GRB\,080916A}& 0.689 &  $$ & $$ & $$ & $$ & $$ & ${\bf 23.80\pm0.09}$ & ${\bf 23.52\pm0.08}$ & ${\bf 22.92\pm0.09}$ & ${\bf 22.82\pm0.12}$ \\
\object{GRB\,081007}& 0.529 &  $$ & $$ & $$ & $$ & $ 24.27\pm0.20$ & ${\bf 24.96\pm0.18}$ & ${\bf 24.79\pm0.20}$ & $$ & ${\bf 23.70\pm0.27}$ \\
\object{GRB\,090424}& 0.544 &  $$ & $$ & $$ & $$ & $$ & $22.97\pm0.17$ & $22.22\pm0.14$ & $21.80\pm 0.13$ & $$ \\
\object{GRB\,091018}& 0.971 &  $$ & $$ & $$ & $$ & $$ & ${\bf 23.42\pm0.07}$ & ${\bf 23.51\pm0.11}$ & ${\bf 22.75\pm0.11}$ & ${\bf 22.44\pm0.14}$ \\
\object{GRB\,091127}& 0.490 &  $$ & $24.16\pm0.18$ & $23.73\pm0.18$ & $$ & $$ & $24.23\pm0.26$ & $23.15\pm0.13$ & $22.86\pm0.10$ & $22.39\pm0.28$ \\
\object{GRB\,100621A}& 0.542 &  $21.82\pm0.10$ & $$ & $$ & $$ & $$ & $21.76\pm0.06$ & $21.41\pm0.06$ & $21.10\pm0.06$ & $21.42\pm0.06$\\
\hline\hline
 \multicolumn{11}{c}{Near-Infrared}  \\
\hline                     
Host galaxy & $z$ &  $Y$ & $J$ & $H$ & $H-F160W$ &$K$ & $IRAC1$ & $IRAC2$&&    \\                
\hline
\object{GRB\,050416}& 0.653 &$$ & ${\bf 22.1\pm0.2}$ & &$$ & $22.21\pm0.20$ & $$ & $$&&\\
\object{GRB\,050525}& 0.606& $$ & $24.4\pm0.3$ & &$$ & $>23.08$ & ${\bf >23.05}$ & ${\bf >23.1}$ &&\\
\object{ GRB\,060614} &0.125 & $$ & $$ & $$ && $22.2\pm0.2$ & ${\bf 22.90\pm0.13}$ & $$&&\\
\object{GRB\,060912} &0.937 & $$ & ${\bf >21.9}$ & &$$ & $21.69\pm0.09$ & ${\bf 21.6\pm0.4}$ & $$ &&\\
\object{GRB\,061021}& 0.346 & $$ & ${\bf >22.4}$ && $$ & ${\bf >22.25}$ & ${\bf >23.7}$ & $$ &&\\
\object{GRB\,071112C}& 0.823 & $$ & $$ & &${\bf 23.8\pm0.06}$ & $$ & ${\bf 23.87\pm0.22}$ & $$ &&\\
\object{GRB\,080319B}& 0.937 & $$ & $$ & $$ & $$ & $$ & $$&&\\
\object{GRB\,080430} &0.767 & $$ & $$ & &${\bf 24.5\pm0.1}$ & $$ & ${\bf >23.39}$ & $$ &&\\
\object{GRB\,080916A}& 0.689 & $$ & ${\bf 22.54\pm0.13}$ & &${\bf 22.76\pm0.03}$ & ${\bf 23.05\pm0.23}$ & ${\bf 22.64\pm0.16}$ & $$ &&\\
\object{GRB\,081007} &0.529 & $$ & ${\bf 23.45\pm0.20}$ & &${\bf 23.61\pm0.25}$ & ${\bf >23.8}$ & ${\bf >23.35}$ & $$ &&\\
\object{GRB\,090424} &0.544 & $$ & $21.58\pm0.11$ & $21.20\pm0.32$ && $$ & ${\bf 21.37\pm0.03}$ & $$ &&\\
\object{GRB\,091018} &0.971 & $$ & ${\bf 21.84\pm0.26}$ & $$ && ${\bf 22.26\pm0.16}$ & ${\bf 22.21\pm0.06}$ & $$ &\\
\object{GRB\,091127}&0.490 & $$ & $$ & &$22.88\pm0.04$ & $$ & ${\bf 23.1\pm 0.12}$ & $$ &&\\
\object{GRB\,100621A}& 0.542 & $21.07\pm0.06$ & $21.41\pm0.06$ & $21.17\pm0.14$& & $21.22\pm0.11$ & $\bf 21.32\pm0.08$ & ${\bf 21.36\pm0.06}$ &&\\  
\hline                  
\end{tabular}
\end{sidewaystable*}

\begin{table}
\caption{Absolute $B$ magnitudes (M$_B$; from the SED fitting), L$_K$, and stellar masses of the LGRB host galaxies of the BAT6 sample at $z<1$.}             
\label{mass}      
\centering  
\small                        
\begin{tabular}{l l cc l}        
\hline\hline                 
Host galaxy & z & M$_B$&Log(L$_K$)& Log(M${_\star}$)  \\    
                   &     &            &[erg\,s$^{-1}$\,Hz$^{-1}$]&   [M${_\odot}$]\\    
\hline                        
\object{GRB\,050416}$^{(a)}$& 0.653 & -19.6&28.6&$9.17_{-0.12}^{+0.12}$\\
\object{GRB\,050525}& 0.606& -15.5&28.0&$8.1_{-0.6}^{+0.6}$\\
\object{ GRB\,060614}$^{(a)}$& 0.125 & -15.7&27.2 &$8.09_{-0.17}^{+0.13}$\\
\object{GRB\,060912}& 0.937 &-20.8& 29.2&$9.23_{-0.07}^{+0.06}$\\
\object{GRB\,061021}& 0.346 & -16.3&27.6 &$8.5_{-0.5}^{+0.5}$\\
\object{GRB\,071112C}& 0.823 & -18.5&28.3&$8.89_{-0.18}^{+0.15}$\\
\object{GRB\,080319B}$^{(b)}$& 0.937 & -17.3&27.4&$8.1_{-0.4}^{+0.4}$\\
\object{GRB\,080430}& 0.767 & -18.5&27.8&$8.15_{-0.20}^{+0.12}$\\
\object{GRB\,080916A}& 0.689 & -19.6&28.4&$8.98_{-0.08}^{+0.07}$\\
\object{GRB\,081007}& 0.529 & -17.6&27.8&$8.78_{-0.45}^{+0.47}$\\
\object{GRB\,090424}& 0.544 & -20.0&28.9&$9.38_{-0.19}^{+0.17}$\\
\object{GRB\,091018}& 0.971 &-20.9&29.0&$9.52_{-0.10}^{+0.08}$\\
\object{GRB\,091127}$^{(c)}$& 0.490 & -18.7&28.1&$8.67_{-0.07}^{+0.07}$\\
\object{GRB\,100621A}$^{(d)}$& 0.542 & -20.6&29.0&$9.04_{-0.05}^{+0.06}$\\
\hline                                   
\end{tabular}
\tablefoot{(a) M${_\star}$ consistent with that reported in \cite{Savaglio2009}; (b): M${_\star}$ consistent with what is reported in \cite{Tanvir2010}; (c): M${_\star}$ consistent with that reported in \cite{Vergani2011}; (d): M${_\star}$ consistent with what is reported in \cite{Kruhler2011}.
}
\end{table}

\section{The host galaxy K luminosities and stellar masses }
\label{SEDsection}
We fitted the spectral energy distribution (SED) of the host galaxies with the code {\it LePhare}\footnote{\url{http://www.cfht.hawaii.edu/~arnouts/LEPHARE/lephare.html}} \citep{Arnouts1999,Ilbert2006}. We used the Stellar Population Synthesis (SPS) package developed by \cite{Bruzual2003}, with the BC03\_m52 library. The metallicity is therefore fixed to Z=0.008 (i.e. $\sim0.4$Z$_\odot$). We assumed an universal IMF from \cite{Chabrier2003} and an exponentially declining star
formation history SFR$\,\propto e^{-t/\tau}$, with a $\tau$ grid of 0.1, 0.3, 1, 2, 3, 5, 10, 15, and 30\,Gyr. Dust extinction was applied to the templates
using the \cite{Calzetti2000} law, with a E$_{(B-V)}$ grid going from 0.0 to 0.9 with a 0.05 step. 
The SED fitting plots and results for each host galaxy are reported in Fig.\,\ref{SED} and Table\,\ref{SEDtab}, respectively. 

Two LGRB hosts (GRB\,080319B and GRB\,050525) do not have NIR detections or stringent upper limits, therefore their K luminosity (or stellar mass) is poorly costrained. We report in Table\,\ref{mass} the K luminosity L$_K$ obtained from the SED fitting for each host galaxy. A determination using a K-corrected flux from the closest band has also been retrieved, giving similar results.

The stellar masses (M$_{\star}$) have been obtained by the SED fitting described above and are reported in Table\,\ref{mass}.  
We verified that varying the input metallicity using the BC03\_m62 library (Z=0.02$\sim$Z$_\odot$), or leaving the metallicity free to vary, the stellar mass values obtained are consistent within errors, not affecting significantly the final distribution. We also considered a Small Magellanic Cloud extinction law \citep{Prevot1984}. Again, the resulting stellar masses are consistent within the uncertainties. 
Only 5 of the 14 host galaxies had the stellar mass determined before (see Table\,\ref{mass}). For all these cases, our results are consistent with the values obtained in previous works.

Star-bursting galaxies have their light dominated by the youngest
stellar populations. In relatively evolved galaxies undergoing a burst
of star formation, this implies that the oldest populations can be
difficult to detect since their light can be largely outshined by
the youngest stars. This can result in large systematic errors that lead to
underestimating their stellar mass when using a single stellar
population SED fitting method (e.g.\,\citealt{Pforr2012}). To
investigate the possible impact of such an uncertainty, we followed
\cite{Papovich2001} and re-fitted each observed SED with the
best-fit template to which an additional stellar population of old
stars was added. For each galaxy, this old population was set as the
SPSs with same parameters that the best-fit SED excepts the age, which
was set to the age of the Universe at the observed redshift. In
principle, this can constrain the maximum contribution of old
populations within the photometric error bars (see \citealt{Papovich2001} for details). We find a negligible contribution to the stellar
mass (i.e. variations much smaller than the statistical uncertainty
associated with the best-fit template) in all cases except GRB\,071112C for which the stellar mass slightly increases to Log\,M$_{\star}=9.1$\,M${_\odot}$. This analysis
confirms that LGRB hosts at $z < 1$ are young star-forming objects,
which are dominated by young stars. To
also obtain strict upper limits on the stellar masses of the hosts, we used
the method by \cite{Bell2003} developed using bright local
galaxies, assuming neither evolution in IMF nor a mass-to-light ratio.

\section{Comparison with star-forming galaxies}

\begin{figure}
   \centering
   \includegraphics[width=\hsize]{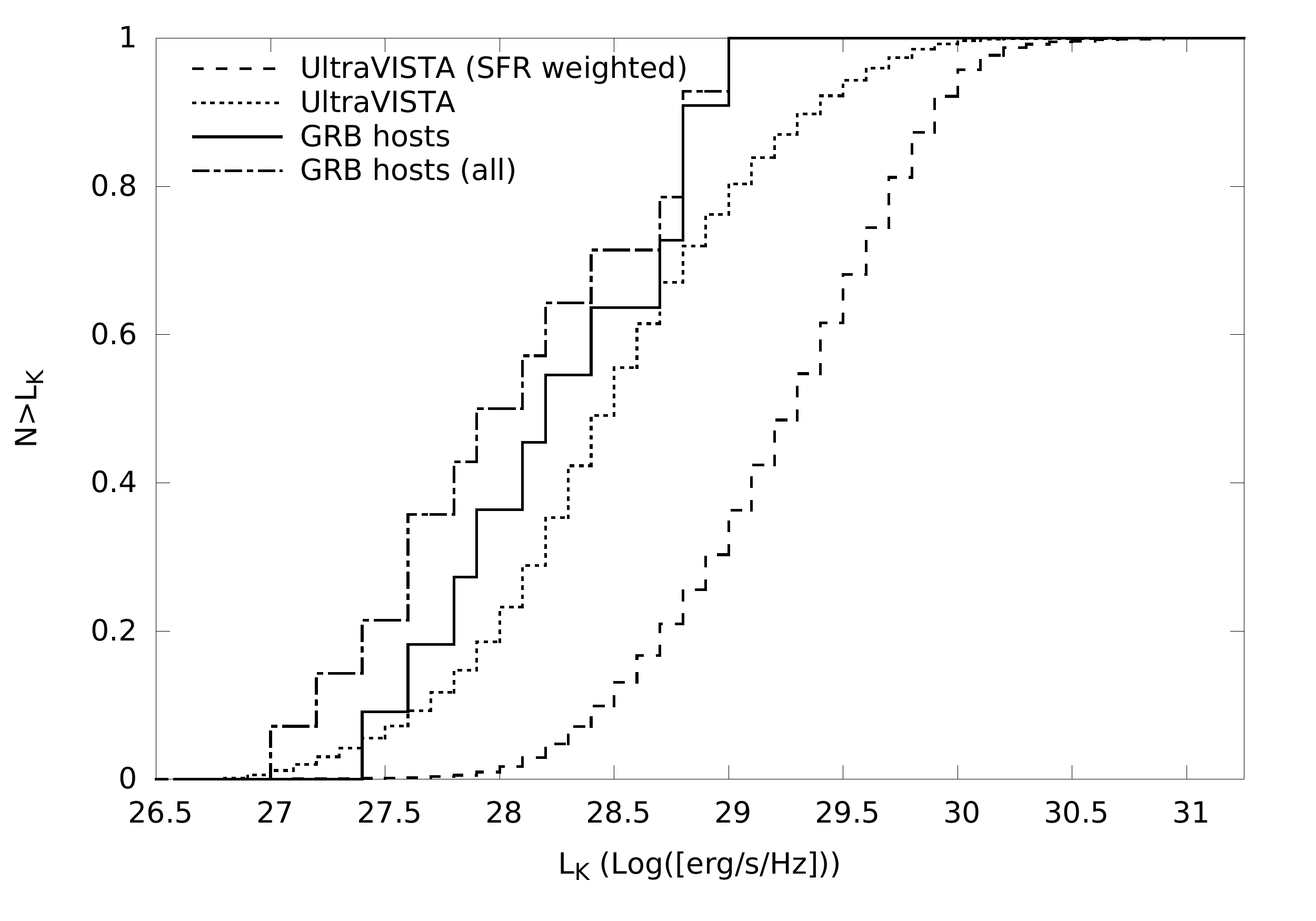}
      \caption{Cumulative distribution of the K luminosity (L$_K$) of the star-forming galaxies in the UltraVISTA survey at $z<1$ (dotted line) weighted by the SFR (dashed line) compared to that of the LGRB hosts in our sample (solid line: without the three LGRB host galaxies not complying with the UltraVISTA redshift and K-band magnitude cuts; dot-dashed line: whole sample).              }
         \label{Klumcum}
   \end{figure}

To investigate the relation between LGRBs and SFR, we compared our LGRB host galaxies to the star-forming galaxies detected the UltraVISTA survey \citep{McCracken2012,Ilbert2013,Muzzin2013}. For this comparison we used the results obtained by the SED fitting described at the beginning of Sect. \ref{SEDsection}, because the SED fitting of the galaxies in the survey (see \citealt{Ilbert2013}) is performed in a similar way.

\subsection{L$_K$ }
\label{LK}

In Fig. \ref{Klumcum} we plot the (normalised) L$_K$ cumulative distribution of the LGRB hosts and of the UltraVISTA star-forming galaxies at $z<1$. We used the UltraVISTA K luminosities and SFRs resulting from the SED fitting obtained by the same authors. The UltraVISTA survey has a cut by definition at m$_K(AB)=24$. The SED fitting of the host galaxies of GRB\,080319B and GRB\,080430 indicates K-band magnitudes lower than this limit. The UltraVISTA results do not consider galaxies at z<0.2. GRB\,060614 lies below this limit. In Fig. \ref{Klumcum} we report the LGRB host L$_K$ cumulative distribution with and without taking the host galaxies into account, and not complying with the survey cuts or having a poor SED (i.e. excluding 3 host galaxies). 

Since LGRBs are the end product of some massive stars, the probability of hosting a LGRB is proportional to the SFR of a galaxy, at first approximation. Therefore, when comparing the hosts with the UltraVISTA galaxies, we should weight the K luminosity of the latter by their SFR. We do this by putting the sum of the SFR values in each bin (resulting from the survey SED fitting) of each galaxy falling in that bin, instead of simply adding the number of galaxies. 

Because of the fast increase in SFR with z, the UltraVISTA SFR-weighted galaxy distribution at $0.2<z<1.0$ might be dominated by galaxies close to $z\sim1$, hence having higher masses due to the UltraVISTA mass completeness limit \citep{Ilbert2013}, whereas the GRB lies at a slightly lower mean redshift ($z=0.64$). This could have some effect on the comparison between the L$_K$ cumulative distribution of the LGRB hosts and of the UltraVISTA star-forming galaxies. Nonetheless, we verified that this effect is small: When considering only the UltraVISTA galaxies at $0.5<z<0.8$ (i.e. around the average redshift of the GRBs in our sample), we obtain a similar SFR-weighted distribution and discrepancy from that of LGRB host galaxies.

Considering L$_K$ as a proxy for the stellar mass, from Fig.\,\ref{Klumcum} we can already infer that the stellar mass distribution of LGRB hosts is not consistent with that of the star-forming galaxies at $z<1$ weighted by they SFR, implying that at $z<1$ LGRBs are not unbiased tracers of the star formation.

\subsection{Stellar mass }
\label{Mstarsection}

\begin{figure}
   \centering
   \includegraphics[width=\hsize]{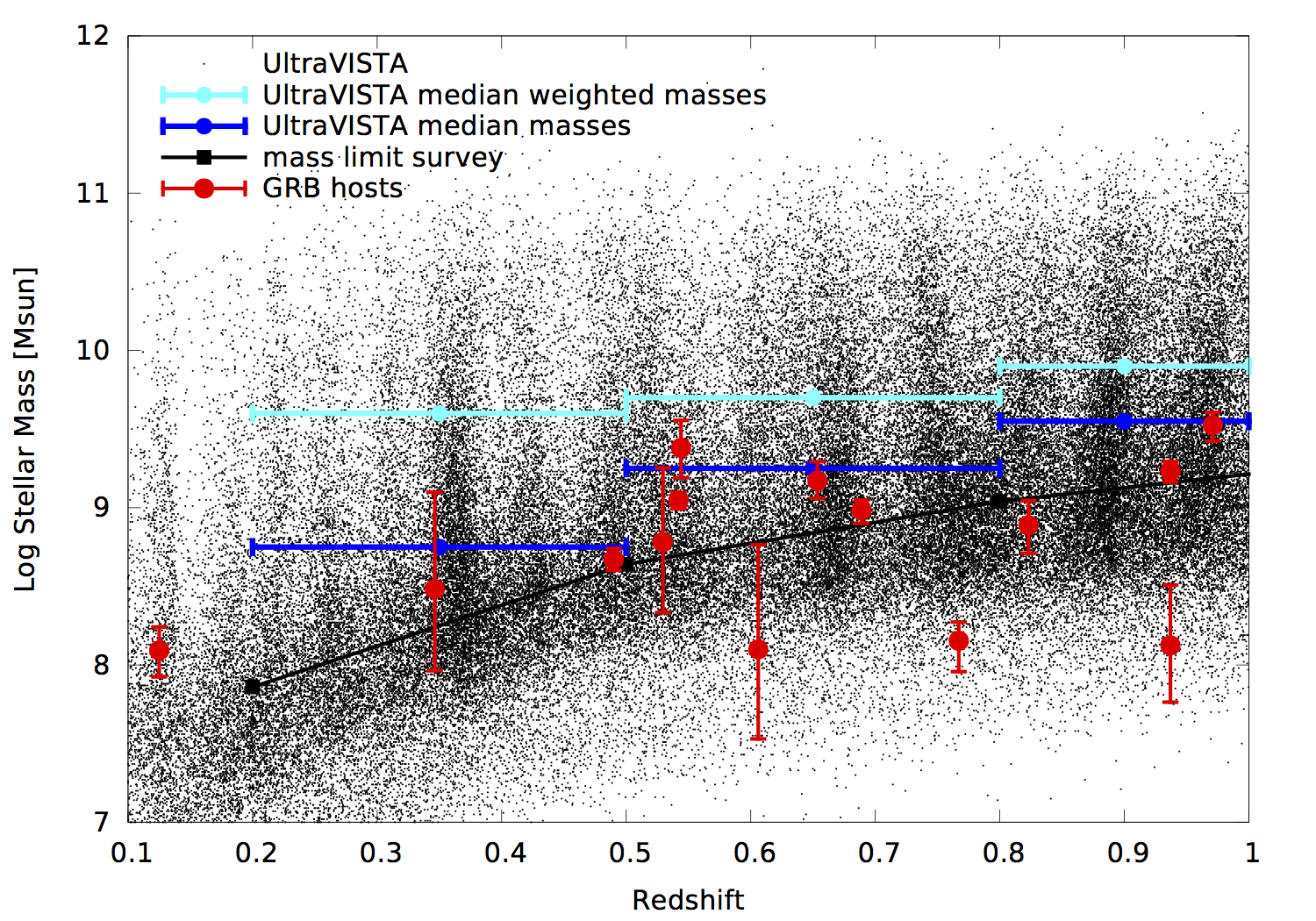}
      \caption{Stellar masses as a function of redshift for the LGRB host galaxies (red and green dots), overplotted on those of the UltraVISTA survey. The black solid line shows the mass completeness of the survey. The blue lines refer to the survey median stellar masses, whereas the cyan lines refer to the same quantity weighted by the galaxy SFR. }
         \label{figmass}
   \end{figure}

In Fig.\,\ref{figmass}, we plot the stellar masses as a function of redshift for the LGRB host galaxies and for the star-forming galaxies of the UltraVISTA survey weighted by their SFR. We considered the mass completeness limits of \cite{Ilbert2013}, also reported in the figure. To weight the galaxy stellar masses by their SFR, 
we proceeded as described in section \ref{LK}. 
As above, we used the SFR values obtained from the SED fitting (SFR$_{SED}$). At this redshift range, the SFR$_{SED}$ is generally found to be statistically in good agreement with the sum of the UV and IR SFR (SFR$_{tot}$; e.g. see \citealt{Wuyts2011}). Furthermore, in our case the use of the SFR$_{tot}$ instead of SFR$_{SED}$ would introduce a bias since the SFR$_{IR}$ is often determined only for massive galaxies.

\begin{figure}
   \centering
   \includegraphics[width=\hsize]{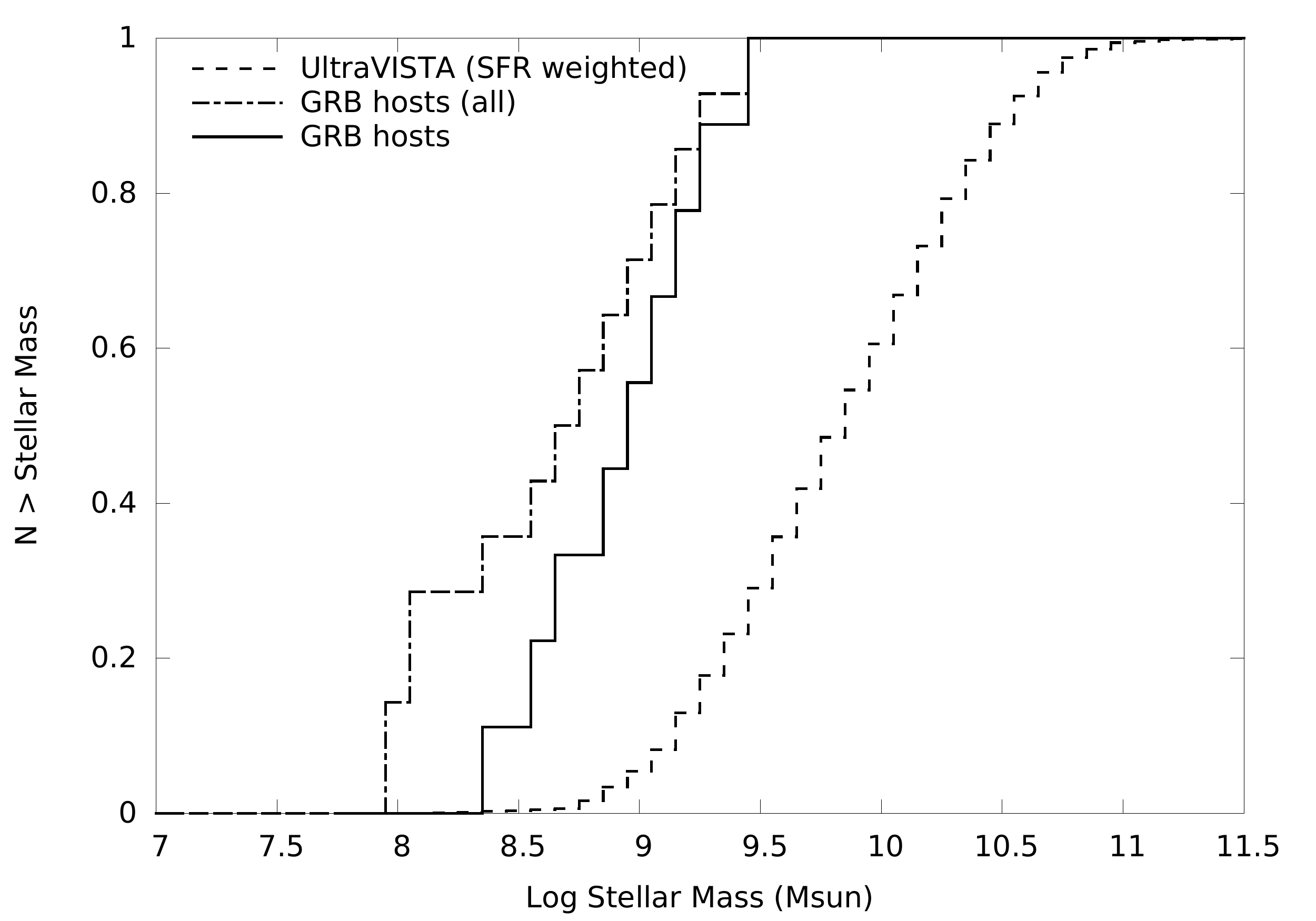}
      \caption{Cumulative stellar mass distribution of the star-forming galaxies in the UltraVISTA survey at $z<1$ weighted by the SFR (dashed line) compared to that of the stellar masses of the LGRB hosts in our sample (dot-dashed line: whole sample; solid line: excluding the five host galaxies not complying with the completeness limits of the UltraVISTA survey).              }
         \label{masscum}
   \end{figure}

In Fig.\,\ref{masscum} we plot the cumulative stellar mass distributions of the LGRB host galaxies and the star-forming galaxies of the UltraVISTA survey at $z<1$ weighted by their SFR, using the same methods as illustrated above. For the comparison with the LGRB host stellar masses, we took the stellar mass completeness of the survey at the different redshifts into
account \citep{Ilbert2013}, as well as the K-magnitude limit. In addition to the three LGRBs excluded for the L$_K$ comparison, we carried out the comparison by not considering the remaining LGRBs having a stellar mass below the completeness limit of the survey. A total of five LGRBs are therefore excluded (the two LGRBs not having detections or deep upper limits in the NIR are among those).  
For the same reason as reported in Sect.\,\ref{LK}, we verified that by
only considering the UltraVISTA galaxies at $0.5<z<0.8$ (always taking the stellar mass completeness of the survey into account) we obtain a similar distribution.

As already suggested by the L$_K$ distributions, the stellar mass distribution of LGRBs is very different from what is expected from the UltraVISTA survey. Indeed, from the Kolmogorov-Smirnov (K-S) test, we can discard the hypothesis that the two samples are drawn from the same distribution ($p=9\times10^{-5}$).

   \section{Boissier's method}
\label{boissier}
Following the method of \cite{Boissier2013}, we can use our sample to determine the LGRB efficiency (the
LGRB rate to SFR rate, called GRB bias in \citealt{Boissier2013}).
This method is based on comparing the stellar masses of
a sample of LGRB hosts and of star-forming galaxies in the same redshift
range (but lower than $\sim1$). In this redshift range, the stellar mass function
is nearly constant in the stellar mass range where it can be measured
(above $\sim10^9$ M$_{\odot}$, e.g. \citealt{Ilbert2013}). 
To compute the efficiency, we use the same
prescription for the star-forming galaxies as in \cite{Boissier2013}. For the
LGRB hosts, we use our sample that has the advantage of being better defined
and complete with respect to the compilation used in that paper.

\begin{figure}
   \centering
   \includegraphics[angle=-90,width=\hsize]{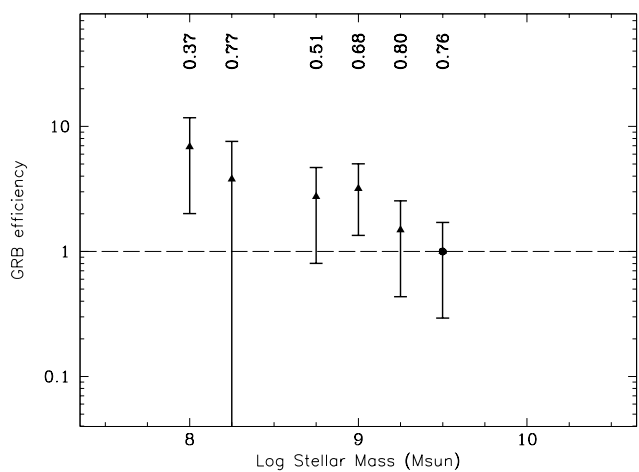}
      \caption{LGRB efficiency (LGRB rate to SFR ratio)
determined from the stellar mass distribution of our sample,
following the method presented in \cite{Boissier2013}.
The results are normalised at log(M)=9.5\,M$_{\odot}$. The
numbers above the points indicate the median redshift of the hosts contributing to
the respective stellar mass bin.              }
         \label{Boissier}
   \end{figure}

This method is in principle very similar to the comparison in the previous section, but includes prescriptions on star-forming galaxies allowing us to compute the variation in the efficiency with stellar mass, taking the median redshift of each bin into
account.
The results (see Fig.\,\ref{Boissier}) have large error bars because of small statistics and are extrapolated
at low masses (assuming the stellar mass function does not vary with redshift also at these
masses); however, we confirm the trend towards decreasing efficiency with stellar mass, already hinted at in
\cite{Boissier2013}.

\section{Inferring the metallicity bias}

We can compare the stellar mass distribution of the LGRB hosts with those found from simulations. We applied the method of \cite{Campisi2009}, using a galaxy catalogue constructed by combining high-resolution N-body simulations with a semi-analytic model of galaxy formation.\footnote{The simulations are a good match to galaxy mass functions \citep{Kitzbichler2007} and reproduce the mass-metallicity relation and the FMR \citep{Campisi2011}.}
The candidate host galaxies of LGRBs are selected by extracting the information for the age and metallicity of newly formed stars. To each galaxy we assigned a probability of hosting a LGRB proportional to the mass of young (i.e. with ages less than a few Myrs) star particles in the simulated galaxy. This is in line with our observational and theoretical understanding of the LGRB events that indicate massive stars are progenitors of LGRBs. We also built different samples by considering different metallicity thresholds for the progenitors as suggested by some theoretical models (see \citealt{Yoon2005,Yoon2006,Woosley2006,Georgy2009}): from no threshold to $Z_{th}=0.1\;Z_\odot$. In these cases, the probability assigned to each galaxy was taken to be proportional to the mass of young star particles with a metallicity below $Z_{th}$.

\begin{figure}[h!]
   \centering
   \includegraphics[width=8cm]{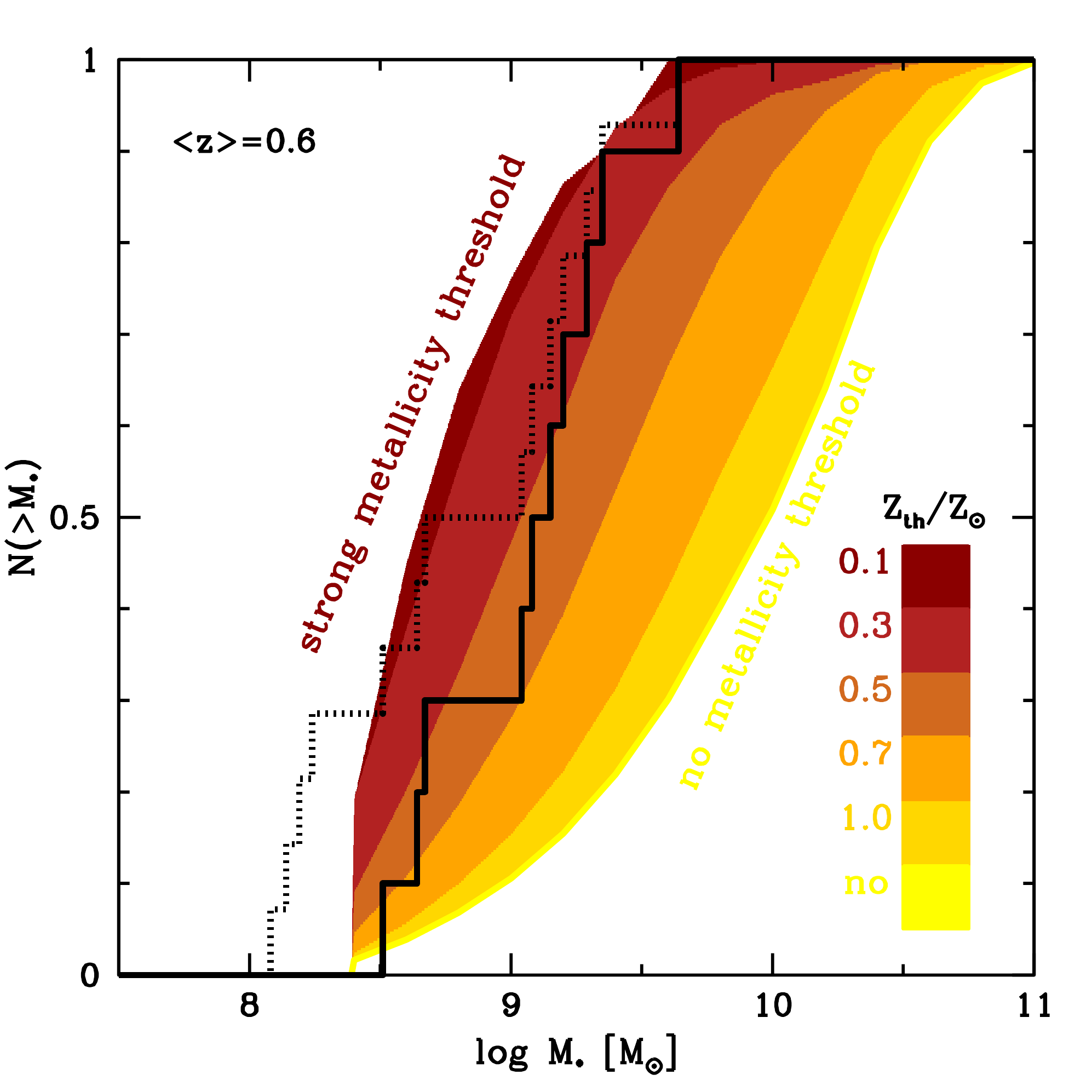}
        \caption{Cumulative distribution of the stellar masses (including the old stellar population) of the LGRB hosts in our sample (dotted line: whole sample; solid line: with LogM$_{\star}$>8.4M$_{\odot}$, to comply with the simulation limits) compared to what is obtained by LGRB host galaxy simulations. The observed distribution is explained by a metallicity cut favouring low metallicities ($Z<0.5Z{_\odot}$). 
                        }
         \label{simul}
   \end{figure}

Since our observed sample is limited to the redshift range $z=0-1$, we have considered here the simulated galaxy population at the mean redshift of $<z>=0.6$. We also tested that the results change only a little by considering a wider redshift range in the simulation within $z<1$. 
For each simulated host galaxy sample we built up the cumulative distribution of stellar masses that we compared with the observed distribution of our complete sample in Fig.\,\ref{simul}.
For this plot, we can use the stellar masses obtained by the SED fitting including the correction for old stellar population (see the end of Sect.\,\ref{SEDsection}). As explained above, such a correction does not change the mass distribution significantly.
From the plot it is clear that the observed distribution is very far from that expected in the case of no metallicity threshold. In fact, the K-S test performed in Section\,\ref{Mstarsection} applies to this case. A strong metallicity threshold, i.e. $Z_{th}=0.1$, predicts a distribution that is too biased towards low stellar masses. The distribution of our complete sample seems to favour a metallicity threshold $Z_{th}=0.3-0.5\;Z_\odot$.

\section{Discussion}

Our results show that LGRB host galaxies at $z<1$ are much fainter than would be expected if GRBs were direct tracers of star formation. This result has also been confirmed by the findings of \cite{Perley2015a}.

\begin{figure}
   \centering
   \includegraphics[width=\hsize]{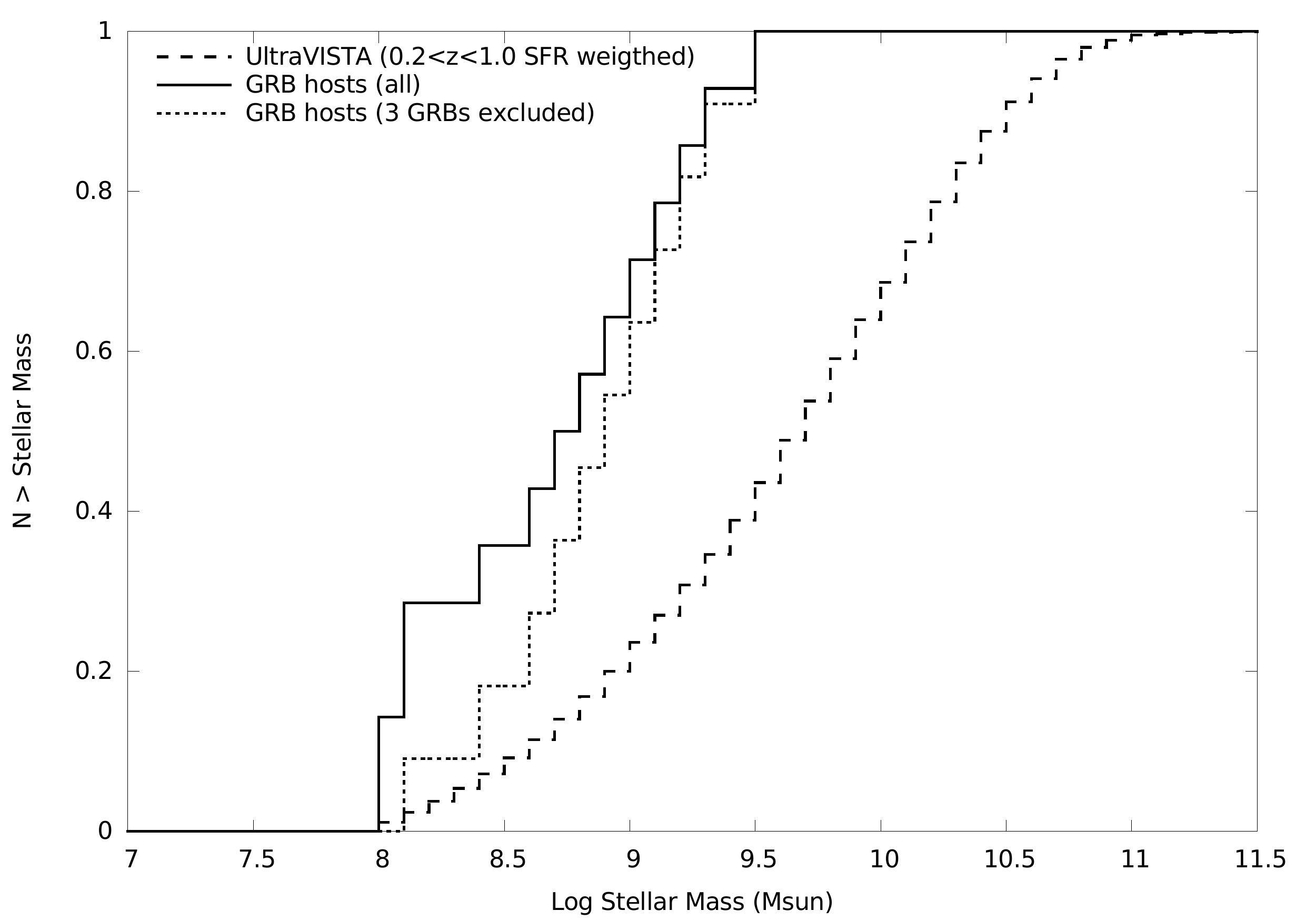}
      \caption{Comparison between the artificially extended stellar mass cumulative distribution of the star-forming galaxies in the UltraVISTA survey at $z<1$, weighted by the SFR, (dashed line) and that of the LGRB hosts in our sample (solid line: whole sample; dotted lines: excluding three GRBs as explained in the text). Illustrated here is the slope of unity case for the SFR-M relation (see text for details).             }
         \label{extended}
   \end{figure}

Some of the GRB host galaxies stellar masses are close to the UltraVISTA completeness limits. To check and strengthen the validity of our results, we can compare the stellar mass distribution of GRB host galaxies to an artificially extended distribution of the UltraVISTA star-forming galaxies down to LogM$_{\star}$=8.0M$_{\odot}$. We proceeded as follows: (i) for each redshift bin, we considered the mass function of star-forming galaxies of \cite{Ilbert2013}, and supposing no evolution of its faint end slope, we extended it to LogM$_{\star}$=8.0M$_{\odot}$; (ii) to weight the SFR, we applied a SFR-M relation to determine the average SFR for each stellar mass bin.
The value of the slope of this relation is still being debated, and it generally varies between a slope of 0.5 and a slope of unity (e.g. \citealt{Wuyts2011} and references therein). We performed our test using a slope of unity but also the more conservative case of a slope of 0.5 (\citealt{Atek2014}, using observations with the HST Wide Field Camera 3, report a slope of 0.65 down to LogM$_{\star}$=8.0M$_{\odot}$). This last case would increase the weight of low-mass galaxies compared to that of a much steeper slope and therefore bring the UltraVISTA distribution closer to that of GRB host galaxies. In both cases the discrepancy between the distribution is still present (see Fig.\,\ref{extended}). A KS test gives a null probability of $p=(0.1,1.7)\times10^{-3}$, respectively, for the two slope values ($p=(0.9,7)\times10^{-3}$, without considering the two GRBs without strong limits in K and the one at $z<0.2$).

To make the two distributions consistent, a steeper faint end of the mass function would be needed, even for the 0.5 slope case. For example, considering the redshift bin $0.5<z<0.8$, the $\alpha_2$ index form Table 2 of \citealt{Ilbert2013} should pass from $\alpha_2=-1.43$ to $\alpha_2=-1.86$ for the half case, and to $\alpha_2<-2.2$ for the unity case (to $\alpha_2<-1.63,<-2.1$, respectively, if the two GRBs without strong limits in near-infrared and the one at $z<0.2$ are excluded). To our knowledge, no survey has ever found such steep slopes at $z<1$, and those values correspond better to the slopes found at $z>2$ (e.g. \citealt{Santini2012}). If this condition is not fulfilled, the consistency can only be achieved if a shallower SFR-M relation describes the low-mass galaxy population. For a flat SFR-M relation, the two distributions would be consistent.

In our $z<1$ sample, {\it \emph{dark}} LGRBs may be slightly underrepresented, even if not statistically significant, with respect to the total sample (20-30\% expected, depending on the definition; Melandri et al. 2012). Under the hypothesis that {\it \emph{dark}} bursts are preferably associated with massive hosts (we note nonetheless that the host of the dark LGRB in our sample is not massive: GRB100621, Log(M${_\star})=9.0 \pm 0.5$ M${_\odot}$; see also \citealt{Perley2013} for further examples), we verify that the addition of a few massive hosts would imply only a small change to the median of the stellar mass distribution, still indicating the preference for low metallicities ($Z<0.5\,Z{_\odot}$). 

We furthermore checked the solidity of this result by using the strict stellar mass upper limits obtained using the \cite{Bell2003} method. The resulting metallicity threshold upper limit is  $Z_{th}=0.7\,Z{_\odot}$.

The redshift completeness of the BAT6 sample is  95\%. Two GRBs of the sample lack redshift information. Again, even under the strong assumption that both of them occurred in massive host galaxies at z < 1, the change in the median of the stellar mass distribution would not be enough to significantly impact our results.

To further quantify the observed discrepancy between the GRB and field galaxy stellar mass distribution, we attempted to also perform the following test.
For each GRB in the sample, we extracted a galaxy from the UltraVISTA star-forming galaxy sample randomly, following the redshift distribution of the GRBs in our sample and taking into account that the probability to find a GRB is proportional to the SFR. 
For each selected galaxy, we picked the corresponding stellar mass values and used them to build their stellar mass distribution. We compared this distribution with that of LGRB host galaxies (with and without taking the mass limits of the survey into
account) using the KS test. We repeated this procedure one thousand times.
We find that in more than 90\% of the cases, we can exclude that the two distributions come from the same parent population. If we consider only the GRB host galaxies complying with the survey completeness limits,
we can rule out the null hypothesis in 80\% of the cases. 
In this last scenario, the only case for which we have a result that does not allow us to discard the inconsistency of the two distribution is under the very unlikely assumption that both the GRBs without redshift information of the BAT6 sample (2 over 58) are at z<1 and occurred in high-mass galaxies. Even if the results support our conclusions, we stress that this test is not appropriate to our small dataset, because the statistics is strongly affected by the fact that we are using a small sample. Solid conclusions based on this test can only be made with much larger complete samples, which are not available today at $z<1$.

Owing to the mass-metallicity relation, low metallicity is expected considering the low masses found for LGRB hosts. The comparison with the simulations allows a more quantitative determination. The metallicity threshold we found does not support a theoretical model that predicts extremely low metallicities for LGRB progenitor stars but agrees with the ones requiring more moderate thresholds (e.g. \citealt{Georgy2009}). 
 
We stress that our result is not at odds with the observation that some GRB host galaxies have metallicities Z>Z$_{th}$. Instead, our result should be interpreted as a preference for
LGRBs at z<1 to explode in low-metallicity galaxies (see e.g. \citealt{Modjaz2008} and \citealt{Kruhler2015} for consistent observational results). Moreover, the Z$_{th}$ of the simulations refers to the progenitor region. 

The moderate metallicity threshold inferred by our study is also predicted by some simulations or semi-analytical models of LGRB hosts. \cite{Trenti2014} consider two channels for the LGRB production: a collapsar (hence requiring metal-poor environments) and a binary system (hence metal independent). Even if their best-fit model includes a moderate metallicity bias, at low redshift it predicts too high an efficiency for the metallicity-independent channel. The authors stress the importance of having a complete sample available to improve their results. Indeed our findings, based on the BAT6 complete sample, confirm the necessity of a higher percentage of collapsar progenitors at $z<1$.

The preference for GRBs to occur in low-mass galaxies has also been found by \cite{Perley2015b,Perley2015a} for a complete sample of radio, {\it Spitzer}-observed GRB host galaxies, respectively. A metallicity dependence of the GRB efficiency would be a natural consequence of the conditions necessary to the progenitor star to produce a GRB. When taking the correlation between stellar mass and metallicity into account, it is not surprising to find that LGRBs tend to explode in low-mass galaxies. What is still not clear is whether metallicity is the only factor affecting GRB efficiency (e.g. variable IMF could also play a role). \cite{Perley2015b} suggest that an enhancement of the GRB efficiency in high specific SFR (sSFR) galaxies could also be present, therefore implying a little dependence on bulk galaxy star formation rate, and not necessarily indicating a dependence on metallicity.

\cite{Salvaterra2012} show that a scenario where GRBs trace the cosmic star formation and their luminosity function is constant with redshift would result in a GRB distribution that peaks at a lower redshift than observed. To recover the observed redshift distribution (keeping the luminosity function constant), cuts at very low metallicity are necessary ($Z_{th}<0.3\,Z{_\odot}$). However, this is contradicted by our results:  $Z_{th}=0.3\,Z{_\odot}$ is really the limit condition for which a luminosity function evolution is not necessary. For a threshold $Z_{th}\sim0.5\,Z{_\odot}$, the typical burst luminosity should increase with redshift as $(1 + z)^{1.3\pm0.6}$ \citep{Salvaterra2012}.

To speculate on the prediction of the epoch at which GRB would become direct tracers of the SFR, we can put our metallicity threshold results in the formula reported by \cite{Langer2006} for the number of stars born with metallicity below a specified value. Assuming that the GRB luminosity function and density do not vary with redshift, and under the hypothesis made by \cite{Langer2006}, to obtain their eq.\,(5), we find that for $Z_{th}$ in the range of 0.3-0.5, the metallicity bias would disappear at $z\sim4$ (see \citealt{Greiner2015,Perley2015a,Kruhler2015} for works on GRB has star formation tracers at $z\ge2$).

\section{Conclusion and perspectives}

We compared the luminosities and stellar masses of the host galaxies of the BAT6 sample at $z<1$ to those of star-forming galaxies in the UltraVISTA survey within the same redshift range. 
We found that at $z<1,$ LGRBs tend to avoid massive galaxies and are very powerful in selecting a population of faint star-forming galaxies, partly below the completeness limits of galaxy surveys. We studied the luminosity and stellar mass distribution and found that LGRB hosts tend to have lower luminosities and stellar masses than what was expected if LGRBs were unbiased star formation tracers, as also found in previous studies on LGRB hosts. To make the two distributions consistent, a much steeper faint end of the mass function would be required, or a shallow SFR-mass relation for the low-mass galaxy population.

The use of the BAT6 complete sample keeps these results from being affected by possible biases that could influence past results based on incomplete samples.

It is important to understand the reasons for this low LGRB efficiency. To investigate, we compared the distribution of the stellar masses of LGRB hosts with the results of simulated LGRB host galaxies with different metallicity cuts on the progenitor star. We showed that the distribution is reproduced well for a metallicity threshold of $Z_{th}=0.3-0.5\;Z_{\odot}$. This can be a consequence of the particular conditions needed for the LGRB progenitor star to produce a LGRB.

The first step for future works will be to determine the metallicities and SFRs of the host galaxies of the BAT6 sample at $z<1$ from their spectra and to compare them with those found for star-forming galaxies in galaxy surveys. Because the sSFR has also been invoked as a possible bias in the LGRB-SFR efficiency, we will perform a similar study on the sSFR of the same sample of galaxy as well. 
Ultimately, we plan to extend these studies to a higher redshift range, in order to determine if and how the LGRB efficiency evolves.

\begin{acknowledgements}
We thank the referee for the constructive suggestions for improving the paper.
We acknowledge Maria Angela Campisi for making her simulations available, Olivier Ilbert and the COSMOS team for helping with the UltraVISTA data and sharing their catalogues, Daniele Malesani and Sandra Savaglio for useful discussions. SDV and ELF acknowledge the UnivEarthS Labex programme at Sorbonne Paris Cit\'e (ANR-10-LABX-0023 and ANR-11-IDEX-0005-02). AFS acknowledges support from grants AYA2010-22111-C03-02 and AYA2013-48623-C2-2 from the Spanish Ministerio de Economia y Competitividad, and grants Prometeo 2009/064 and PrometeoII 2014/060 from the Generalitat Valenciana. This work is partially based on observations with: the ESO Telescopes at Paranal Observatory under programme ID 092.D-0305, PI: S.D. Vergani and ID 092.A-0231, PI: T. Kr\"uhler; the Gemini Observatory, acquired through the Gemini Science Archive, which is operated by the Association of 
Universities for Research in Astronomy, Inc., under a cooperative agreement with 
the NSF on behalf of the Gemini partnership: the National Science Foundation 
(United States), the National Research Council (Canada), CONICYT (Chile), the 
Australian Research Council (Australia), Minist\'erio da Ci\^encia, Tecnologia e 
Inova\c c\'ao (Brazil) and Ministerio de Ciencia, Tecnolog\'a e Innovaci\'on Productiva 
(Argentina).; GROND (\url{http://www.mpe.mpg.de/~jcg/GROND/}); Gran
Telescopio Canarias (GTC), instaled in the Spanish Observatorio del Roque
de los Muchachos of the Instituto de Astrof\'isica de Canarias in the island of
La Palma;  the Nordic Optical Telescope, operated
on the island of La Palma jointly by Denmark, Finland, Iceland, Norway,
and Sweden, in the Spanish Observatorio del Roque de los Muchachos of the Instituto de Astrof\'isica de Canarias; 
the NASA/ESA Hubble Space Telescope, obtained from the data archive at the Space Telescope Science Institute. STScI is operated by the Association of Universities for Research in Astronomy, Inc. under NASA contract NAS 5-26555;
the SDSS (Funding for SDSS-III has been provided by the Alfred P. Sloan Foundation, the Participating Institutions, the National Science Foundation, and the U.S. Department of Energy Office of Science. The SDSS-III web site is http://www.sdss3.org/. SDSS-III is managed by the Astrophysical Research Consortium for the Participating Institutions of the SDSS-III Collaboration including the University of Arizona, the Brazilian Participation Group, Brookhaven National Laboratory, Carnegie Mellon University, University of Florida, the French Participation Group, the German Participation Group, Harvard University, the Instituto de Astrofisica de Canarias, the Michigan State/Notre Dame/JINA Participation Group, Johns Hopkins University, Lawrence Berkeley National Laboratory, Max Planck Institute for Astrophysics, Max Planck Institute for Extraterrestrial Physics, New Mexico State University, New York University, Ohio State University, Pennsylvania State University, University of Portsmouth, Princeton University, the Spanish Participation Group, University of Tokyo, University of Utah, Vanderbilt University, University of Virginia, University of Washington, and Yale University); the Spitzer Space Telescope, which is operated by the Jet Propulsion Laboratory, California Institute of Technology under a contract with NASA; the Italian Telescopio Nazionale Galileo (TNG) operated on the island of La Palma by the Fundaci\'on Galileo Galilei of the INAF (Istituto Nazionale di Astrofisica) at the Spanish Observatorio del Roque de los Muchachos of the Instituto de Astrofisica de Canarias". 
Part of the funding for GROND (both hardware as well as personnel) was generously granted from the Leibniz-Prize to Prof. G. Hasinger (DFG grant HA 1850/28-1).
Part of the research leading to these results has received funding from the European Community's Seventh Framework Programme (FP7/2013-2016) under grant agreement number 312430 (OPTICON).
This research has made use of the NASA/ IPAC Infrared Science Archive, which is operated by the Jet Propulsion Laboratory, California Institute of Technology, under contract with the National Aeronautics and Space Administration. This research has made use of the \url{http://www.mpe.mpg.de/~jcg/grbgen.html} page and of the GHostS database (www.grbhosts.org), which is partly funded by Spitzer/NASA grant RSA Agreement No. 1287913.

      \end{acknowledgements}

\bibliographystyle{aa} 
\bibliography{susy14} 

\Online

\begin{appendix} 
\label{appendix}
\section{SED data and plots}
\begin{figure*}
\centering
\includegraphics[width=5cm]{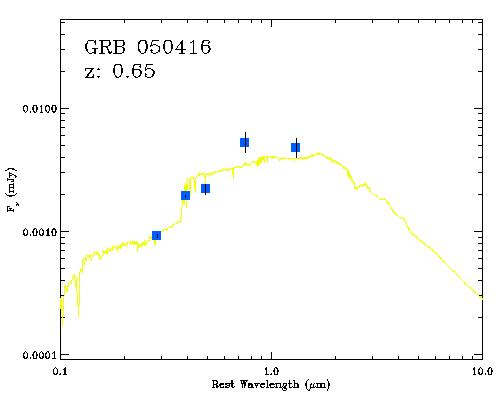}
\includegraphics[width=5cm]{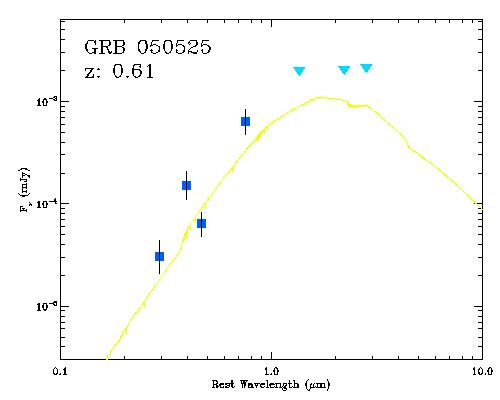}
\includegraphics[width=5cm]{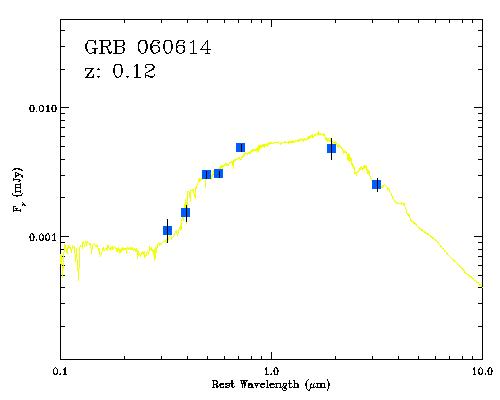}
\includegraphics[width=5cm]{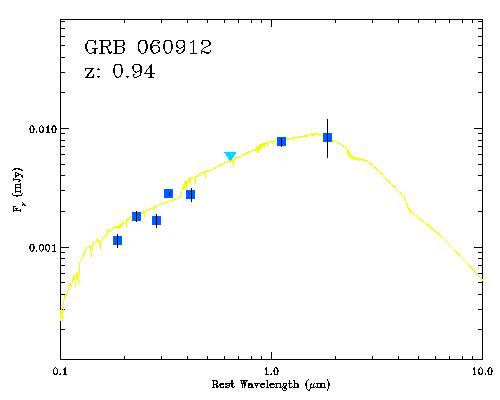}
\includegraphics[width=5cm]{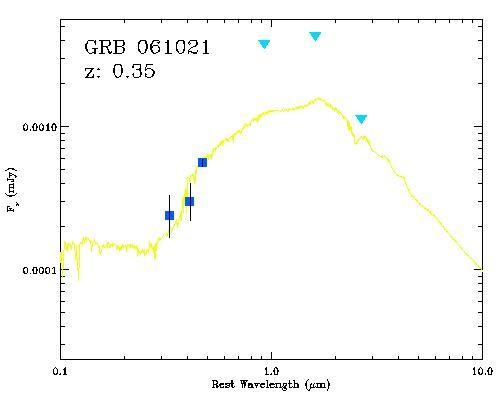}
\includegraphics[width=5cm]{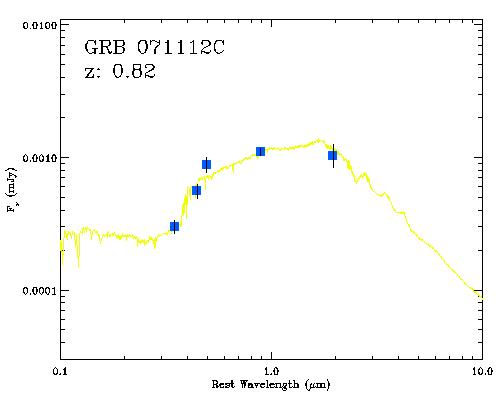}
\includegraphics[width=5cm]{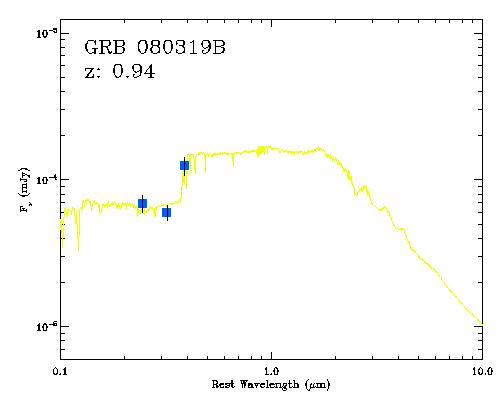}
\includegraphics[width=5cm]{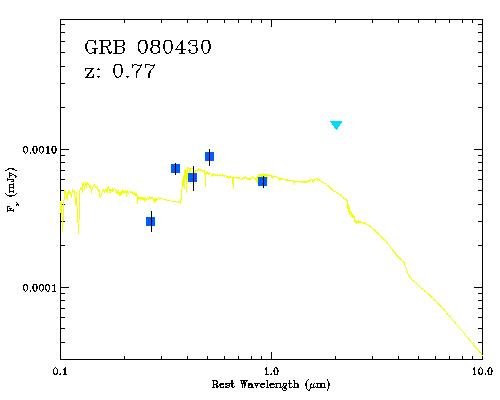}
\includegraphics[width=5cm]{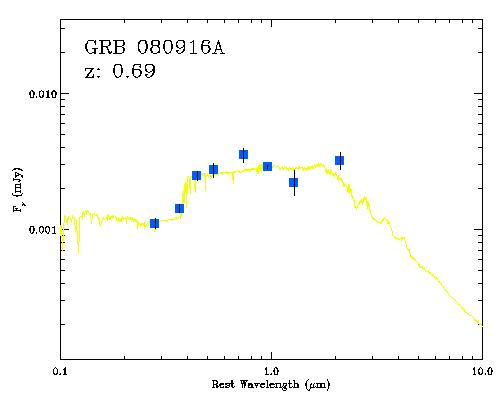}
\includegraphics[width=5cm]{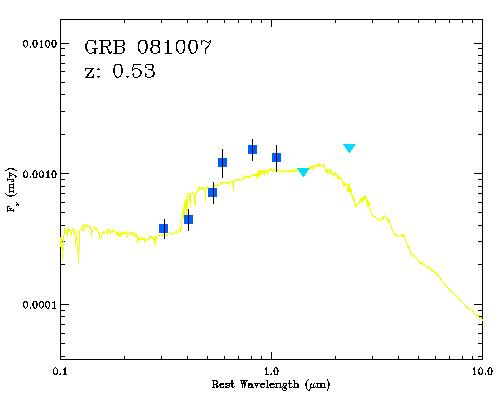}
\includegraphics[width=5cm]{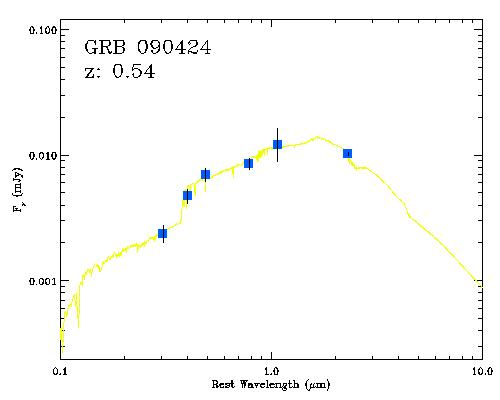}
\includegraphics[width=5cm]{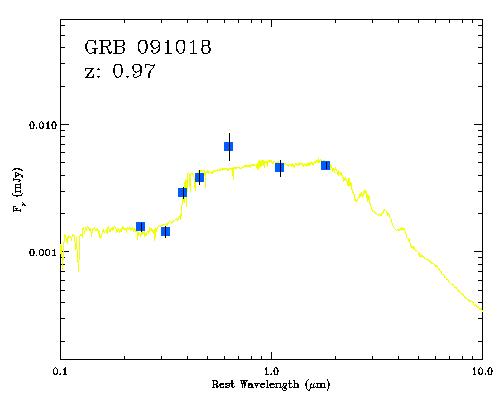}
\includegraphics[width=5cm]{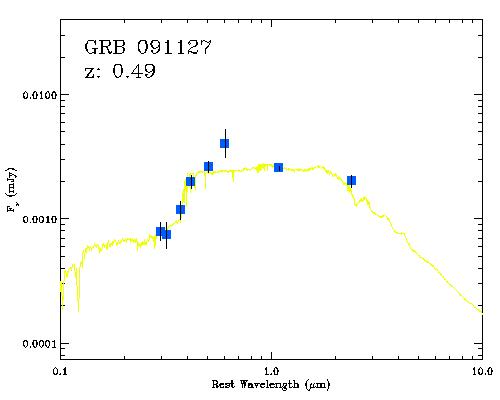}
\includegraphics[width=5cm]{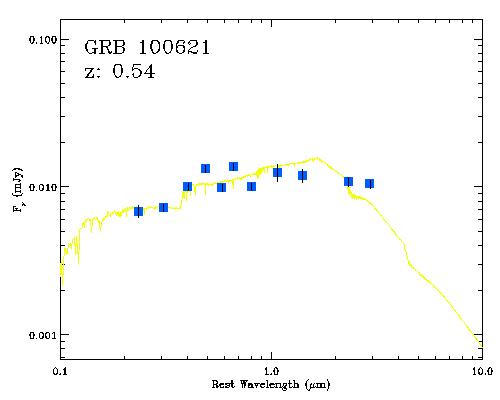}

\caption{SED plots}
\label{SED}
\end{figure*}

For each host galaxy we report the data used with a short description of the newly obtained ones.
The SED fitting plots and results for each host galaxy are reported in Fig.\,\ref{SED} and Table\,\ref{SEDtab}, respectively.


\subsection{GRB\,050416A host}
To build the SED of this host we used the public TNG/DOLORES $J$-band data (\url{http://ia2.oats.inaf.it/index.php/tngarchive/tng}; programme ID: A22\_TAC107 PI:D.Malesani) and the photometry published by \cite{Soderberg2007,Perley2009,Hjorth2012}.

\subsection{GRB\,050525A host}
To build the SED of this host, we obtained GTC/OSIRIS data with the $g i$ filters (programme ID: GTC31-13B PI: A. Fernandez-Soto), and we used the $R$ and $K$-band photometry published by \cite{Hjorth2012},
as well as the public {\it Spitzer} IRAC1and IRAC2 data obtained under programme ID:\,3653, PI: P.M. Garnavich.
The GTC/OSIRIS observations were obtained on the October 23, 2013 for a total exposure of $6\times300$\,s and $15\times160$\,s in {\it g} and {\it i} bands, respectively.

\subsection{GRB\,060614 host}
To build the SED of this host, we used the photometry published by \cite{Della-Valle2006,Mangano2007a,Hjorth2012} and the public {\it Spitzer} IRAC1 data obtained under programme ID:\,90062, PI: D.A. Perley.

\subsection{GRB\,060912A host}
To build the SED of this host, we used the public TNG/DOLORES $U B V I J$-band data (\url{http://ia2.oats.inaf.it/index.php/tngarchive/tng}; program ID: A22\_TAC107 PI:D.Malesani), the $R$ and $K$ band photometry published by \cite{Hjorth2012}, and the public {\it Spitzer} IRAC1 data obtained under programme ID:\,90062, PI: D.A. Perley.

\subsection{GRB\,061021 host}
To build the SED of this host, we obtained VLT/HAWK-I photometry (ID 092.D-0305, PI: S.D. Vergani). We also used the public TNG/DOLORES $B$ and $V$ band data (\url{http://ia2.oats.inaf.it/index.php/tngarchive/tng}; program ID: A22\_TAC107 PI:D.Malesani), the public {\it Spitzer} IRAC1 data obtained under programme ID:\,90062, PI: D.A. Perley, and the $R$-band photometry published by \cite{Hjorth2012}.

VLT/HAWK-I observations were taken between December 28, 2013 and January 7, 2014 for a total of 30\,minutes ($15\times120$\,s) in J band and 80\,minutes ($40\times120$\,s) in K band.

\subsection{GRB\,071112C host}
To build the SED of this host we obtained GTC/OSIRIS data with the $z$ filter (programme ID: GTC31-13B PI: A. Fernandez-Soto). We also used $R$-band GEMINI public data (programme ID:\,GN-2008B-Q-112 PI:\,A. Levan), TNG/DOLORES $I$-band data (programme ID: A17\_TAC9 PI:C. Guidorzi), HST F160W data (programme ID:\,GN-2008B-Q-112 PI:\,A. Levan), and the public {\it Spitzer} IRAC1 data obtained under programme ID:\,90062, PI: D.A. Perley.

The GTC/OSIRIS observations were obtained on February 1, 2014 under clear sky conditions for a total exposure of $21\times100$\,s in $z$ band.

\subsection{GRB0\,80319B host}
To build the SED of this host we used the photometry published by \cite{Tanvir2010}.

\subsection{GRB\,080430 host}
To build the SED of this host, we obtained TNG/DOLORES photometry with the $g, i, z$ filters under the OPTICON (Optical Infrared Co-ordination Network for astronomy) programme OPT13B\_71 (PI S.D. Vergani). We also used the r-band Keck LRIS photometry kindly made available by D.A. Perley in his preliminary LGRB hosts catalogue (\url{http://www.astro.caltech.edu/LGRBhosts/}), the HST F160W data (programme ID:\,GN-2008B-Q-112 PI:\,A. Levan), as well as the public {\it Spitzer} IRAC1 data obtained under programme ID:\,90062, PI: D.A. Perley.

The TNG observations were obtained on January 28, 2014 and February 1, 2014 for a total of 8250\,s, 6900\,s, and 9630\,s, in the $g, i, z$ filters, respectively.

\subsection{GRB\,080916A host}
To build the SED of this host, we obtained VLT/HAWK-I photometry (ID 092.D-0305, PI: S.D. Vergani), we used photometry obtained by the GROND (see\,\ref{GROND}), the HST F160W data (programme ID:\,GN-2008B-Q-112 PI:\,A. Levan), and the public {\it Spitzer} IRAC1 data obtained under programme ID:\,90062, PI: D.A. Perley.

The K,J-band VLT/HAWK-I observations started on November 10 and 13, 2013, respectively, under excellent seeing conditions. We obtained a total of 26\,minutes ($13\times120$\,s) in the J band and 25\,minutes ($15\times100$\,s) in the Ks band.
GROND observations started on December 21, 2008. The integration of the individual 12 {\it griz} and 360 {\it JHK} images is of  369\,s and 10\,s, respectively.

\subsection{GRB\,081007 host}
To build the SED of this host, we used VLT/HAWK-I photometry (ID 092.A-0231, PI: T. Kr\"uhler), GROND photometry (see\,\ref{GROND}), the HST F160W data (programme ID:\,GN-2008B-Q-112 PI:\,A. Levan),  the photometry published in \cite{Jin2013} and the public {\it Spitzer} IRAC1 data obtained under programme ID:\,80153, PI: D.A. Perley.

The HST images reveal two objects: one falling in the UVOT afterglow error circle and another separated by less than 1\arcsec ($<6.3$\,kpc). Those are not resolved by the other instruments used, and together they extend over $\sim1.5\arcsec$. Even if it is not certain, in this paper we considered them as forming a single system.

VLT/HAWK-I observations started on October 13, 2013 for a total of 18\,minutes ($36\times30$\,s) in J band and 28\,minutes ($168\times10$\,s) in K band.
GROND observations started on December 22, 2011. The integration of the individual 12 {\it griz} and 360 {\it JHK} images is  375\,s and 10\,s, respectively.

\subsection{GRB\,090924 host}
To build the SED of this host, we used the photometry published in \cite{Jin2013}, that available from the SDSS (\url{https://www.sdss3.org/dr10/imaging/}) and the public {\it Spitzer} IRAC1 data obtained under programme ID:\,90062, PI: D.A. Perley.

\subsection{GRB\,091018 host}
To build the SED of this host, we obtained VLT/HAWK-I photometry (ID 092.D-0305, PI: S.D. Vergani), we used the photometry obtained by the GROND (see\,\ref{GROND}) and the public {\it Spitzer} IRAC1 data obtained under programme ID:\,90062, PI: D.A. Perley.

VLT/HAWK-I observations started on November 13, 2013 under excellent seeing conditions for a total of 40\,minutes ($20\times120$\,s) in K-band.
GROND observations started on August 30, 2011. The integration of the individual 16 {\it griz} and 480 {\it JHK} images is  375\,s and 10\,s, respectively.

\subsection{GRB\,091127 host}
To build the SED of this host, we used the photometry published in \cite{Vergani2011} and the public {\it Spitzer} IRAC1 data obtained under programme ID:\,90062, PI: D.A. Perley.

\subsection{GRB\,100621A host}
To build the SED of this host, we used the photometry published in \cite{Kruhler2011} and the public {\it Spitzer} IRAC1and IRAC2 data obtained under programme ID:\,70036, PI: D.A. Perley.

\begin{table*}
\caption{SED fitting results}             
\label{SEDtab} 
\tiny     
\centering          
\begin{tabular}{l l c cc c c crrr }     
\hline\hline       
Host galaxy      &  z     & $\chi^2$& E$_{(B-V)best}$ &  Log(M$_{\star}$)  &     Log(M$_{\star inf}$)&      Log(M$_{\star sup}$)      & Age     &   Log(SFR)     &  Log(SFR$_{inf}$)     & Log(SFR$_{sup}$) \\
                          &         &               &              & [M${_\odot}$]  &        [M${_\odot}$]  &                   [M${_\odot}$]            & [yr]         &         [M${_\odot}$\,yr$^{-1}$]              &     [M${_\odot}$\,yr$^{-1}$]              &   [M${_\odot}$\,yr$^{-1}$]                \\
\hline
GRB\,050416  & 0.653  & 8.3 & 0.100 & 0.910E+01 &0.901E+01 &0.923E+01 &0.32E+09 &-0.777E-01&-0.418E+00& 0.334E+00   \\
GRB\,050525  & 0.606  & 2.8 &0.900 & 0.810E+01 &0.753E+01 &0.876E+01 &0.10E+08 &-0.284E+00&-0.107E+01& 0.436E+00   \\
GRB\,060614  & 0.125  & 2.6 &0.000 & 0.809E+01 &0.792E+01 &0.824E+01 &0.50E+10 &-0.153E+01&-0.172E+01&-0.130E+01   \\
GRB\,060912  & 0.937  & 6.8 &0.000 & 0.929E+01 &0.917E+01 &0.953E+01 &0.14E+10 & 0.144E+01& 0.997E+00& 0.166E+01   \\
GRB\,061021  & 0.346  & 0.4 &0.000 & 0.848E+01 &0.796E+01 &0.910E+01 &0.90E+10 &-0.391E+00&-0.110E+01& 0.422E+00   \\
GRB\,071112  & 0.823  & 1.8 &0.000 & 0.889E+01 &0.871E+01 &0.904E+01 &0.55E+10 &-0.125E+00&-0.465E+00& 0.336E+00   \\
GRB\,080319  & 0.937  & undef &0.000 & 0.812E+01 &0.776E+01 &0.851E+01 &0.57E+09 &-0.516E+00&-0.881E+00& 0.263E-01   \\
GRB\,080430  & 0.767  & 9.7& 0.000 & 0.815E+01 &0.796E+01 &0.827E+01 &0.18E+09 &-0.580E-01&-0.281E+00& 0.373E+00   \\
GRB\,080916  & 0.689  & 1.9 &0.000 & 0.898E+01 &0.890E+01 &0.905E+01 &0.10E+10 & 0.181E+00&-0.123E-01& 0.372E+00   \\
GRB\,081007  & 0.529  & 2.0 &0.000 & 0.878E+01 &0.833E+01 &0.925E+01 &0.26E+10 & 0.149E+00&-0.530E+00& 0.753E+00   \\
GRB\,090424  & 0.544  & 0.3& 0.250 & 0.938E+01 &0.919E+01 &0.955E+01 &0.18E+09 & 0.913E+00& 0.555E+00& 0.126E+01   \\
GRB\,091018  & 0.971  & 1.6 &0.000 & 0.952E+01 &0.942E+01 &0.960E+01 &0.72E+09 & 0.638E+00& 0.413E+00& 0.882E+00   \\
GRB\,091127  & 0.490  & 1.3 &0.000 & 0.867E+01 &0.860E+01 &0.874E+01 &0.40E+09 &-0.421E+00&-0.847E+00&-0.962E-01   \\
GRB\,100621  & 0.542  & 8.8 &0.200 & 0.904E+01 &0.899E+01 &0.910E+01 &0.40E+08 & 0.141E+01& 0.133E+01& 0.150E+01 \\
\hline                  
\end{tabular}
\end{table*}

\end{appendix}


\end{document}